\documentclass[useAMS,usenatbib]{mn2e}
\usepackage{graphicx}

\def\Msun{\>{\rm M_{\odot}}}

\newcommand{\gtsim}{\mathrel{\hbox{\rlap{\lower.55ex \hbox {$\sim$}}
                   \kern-.3em \raise.4ex \hbox{$>$}}}}
\newcommand{\ltsim}{\mathrel{\hbox{\rlap{\lower.55ex \hbox {$\sim$}}
                   \kern-.3em \raise.4ex \hbox{$<$}}}}

\newcommand{\myedit}{}   

\title{The Lives of High Redshift Mergers}

\author[McCavana {\it et al.}]{Tom McCavana$^{1}$\thanks{E-mail:
mccavana@physics.usyd.edu.au}, Miroslav Micic${^{1,2}}$, Geraint F. Lewis${^1}$, Manodeep Sinha${^3}$,\vspace{2.5mm}\\
\hspace{-1mm}{\LARGE {\rm Sanjib Sharma${^1}$, Kelly Holley-Bockelmann${^3}$ and Joss Bland-Hawthorn${^1}$}}\\
$^{1}$Sydney Institute for Astronomy, School of Physics, A28, The University of Sydney, NSW 2006, Australia\\
$^{2}$Astronomical Observatory Belgrade,Volgina 7, P.O.Box 74 11060 Belgrade, Serbia\\
$^{3}$Department of Physics \& Astronomy, Vanderbilt University, Nashville, TN, USA\\
}

\begin{document}

\maketitle

\begin{abstract}
{\myedit We present a comparative study of recent works on merger-timescales with
 dynamical friction and find a strong contrast between idealized/isolated mergers \citep{boylan}
 and mergers from a cosmological volume \citep{jiang08}. Our study measures the duration of mergers
in a cosmological N-body simulation of dark matter, with emphasis on higher redshifts (z$\leq$10) and
a lower mass range. In our analysis we consider and compare two merger definitions; tidal disruption 
and coalescence. We find that the merger-time formula proposed by \citet{jiang08} describes our results well
and conclude that cosmologically motivated merger-time formulae provide a more versatile and statistically robust
 approximation for practical applications such as semi-analytic/hybrid models.}
\end{abstract}


\section{INTRODUCTION}

Galaxy mergers are aided by dynamical friction, a process that extracts energy and angular momentum from the orbit
of an incoming galaxy through a gravitational wake that acts as a drag force. Its theoretical conception \citep{chandra43}
was based on a point mass traveling through an in-finite, uniform and non-self gravitating medium. With relatively 
few modifications to account for mass loss in the satellite galaxy, it is used to estimate the duration of galaxy
mergers. As such it is a centrepiece of galaxy evolution; for example, in semi-analytical and hybrid models of galaxy
evolution the accuracy of analytical merger-times is directly responsible for quantities such as stellar mass, gas available
to the central black holes, the distribution of galaxy colour and 
morphology (e.g. \citealt{kauffmann93,volonteri03,croton06,miro07,somerville08,miro11}).

In current $\Lambda$CDM cosmology the first objects to form in the universe were made out of dark matter \citep{zwicky33,rubin80}.
 Widely known as Dark Matter Halos (DMHs hereafter), their formation through violent collapse
and relaxation results in a shallow inner density profile, with the remainder of accreted dark matter trailing behind in
a steep outer slope \citep{ciardi+ferrara05}. The first galaxies, known as proto-galaxies, are born at the heart
of these DMHs from gas that has been virialized, radiatively cooled and collapsed into a centrifugally supported
disk. These proto-galaxies are the birth place of quasars and AGNs which later evolve into the spiral and
 elliptical galaxies observed today. As dark matter halos merge into larger structures so do the central proto-galaxies.
 This process (thought to be responsible for the AGN duty cycle) feeds the central AGN by providing a reservoir of gas which
is then accreted onto the central massive black hole. 

The dynamics of DMH mergers is described with reasonable success by a combination of gravitational free-fall,
dynamical friction and tidal stripping. Chandrasekhar’s dynamical friction has been analytically developed to predict
the coalescence time for a circular orbit from some initial radius \citep{binney+tremaine}. Most versions of the 
approximating formula follow the form:

\begin{equation}
T_{merge}=C_{df}\frac{(M_h/M_s)}{\rm ln\ \Lambda}\frac{f(\epsilon)r_*}{V_c(r_{vir})},
\label{eq:generic}
\end{equation}

\noindent where $C_{df}$ is a constant, $M_h$ and $M_s$ are the masses of the host and satellite DMHs
respectfully and $f(\epsilon)$ is a function of circularity introduced analytically by \citet{laceycole}. 
$\rm ln \Lambda$ is the Coulomb logarithm (discussed below),  $V_c(r_{vir})$ is the
 circular velocity at the virial radius ($r_{vir}$) of the host halo, and $r_*$ is a radius that varies
depending on assumptions made. For bound circular orbits, $r_*$ is the initial radius $r_i$ \citep{binney+tremaine}.
In the more general case of a bound orbit (radial or tangential) $r_*$ is $r_c(E)$ \citep{laceycole}. $r_c(E)$ is the radius that a
circular orbit would have with the same orbital energy. $r_*$ is also often approximated by $r_{vir}$ as most works
  assume $r_i \approx r_{vir}$ or $r_c(E)/r_{vir} \approx 1$.

Studies to improve the approximation of DMH merger-times (and by proxy the effects of dynamical friction) are
numerous and range from analytical/semi-analytical
(e.g.\citealt{weinberg89,tremaine+weinberg,laceycole,benson05a,taylor+babul01,vandenbosch99,colpi}) to numerical 
(e.g. \citealt{ahmad+cohen,lin+tremaine,tormen98,hashimoto03,fujii06}). Significant
work has also been done to assess the accuracy of the Coulomb logarithm (see e.g. Jiang et al. 2008). Analytically, 
$\rm \Lambda$ is found to be a function of the maximum and minimum impact parameters. For extended bodies a better agreement is
found with $\rm ln \Lambda=\rm ln (1+M_h/M_s)$, where the satellite mass is determined at in-fall.

{\myedit In order to refine the accuracy of dynamical friction formulae, Boylan-Kolchin et al. (2008, BK08 hereafter)
 explore merger timescales over a range of mass ratios, circularities and orbital energies using a suite of simulations
 of isolated mergers. In their analysis they build two well resolved Hernquist halos and merge them in various 
 combinations of initial orbital parameters. They also explore the influence of baryons and find that including a stellar bulge
 ($M_*/M_{DM}=0.05$) shortens the merger-time by $\approx10\%$. BK08 produce a empirical formula that relates
 merger-time to a mergers initial orbital parameters. They argue that a typical application of equation \ref{eq:generic}
systematically underestimates the merger-times seen in their simulation. This discrepancy grows as the host becomes 
 larger than the satellite. 

As one of the first simulations to study dynamical friction in a cosmological context, Jiang et al. (2008, J08
 hereafter) developed a merger-time formula using direct measurements from a high resolution cosmological SPH/N-body 
simulation. Their measurements provide a more physically motivated approach for estimating merger-times. Galaxies merge in
the astrophysically realistic context of cosmological structure formation. In their analysis they take the analytical formula of
 \citet{laceycole} and break it down into its principle components, fitting each parameter in order of importance using
their measurements of mergers between 0$\leq$z$\leq$2. J08 find that using alternate forms of Coulombs logarithm gives a
 better agreement in the mass ratio dependence. They also propose a new form of the circularity dependence that better
 accounts for radial orbits (where the formerly used powerlaw breaks down). In contrast to \citet{nfw95}, J08 find that
 the commonly used formula from \citet{laceycole} systematically underestimates the merger-time for minor mergers and 
overestimates the merger-time for major mergers. They conclude their study by providing a statistically robust and
 cosmologically founded formula for merger-times. A follow up work \citet{Jiang10} addresses minor issues in the 
baryonic physics such as overcooling. 

In the current work we extend J08s study in two ways. Firstly by confirming their result using a different area of
 structural evolution, i.e. looking at higher redshifts (z$\leq$10) and a lower mass range. 
To this end we find our results are described well by J08s fitting formula. 
 The second is by giving an independent comparison with BK08 who published
 at around the same time. Both works look at the impact of orbital parameters on merger-timescales, however in very 
different ways. BK08 are able to make clean and accurate measurements as they are dealing with an isolated/closed 
system. J08 conversly, are looking at mergers amidst cosmological structure formation where kinematics can be affected 
by the local enviroment. 

One difference between J08 and this study is the way in which merger-time is defined. We explore
 two types of merger; mergers that finish with coalescence of the two halo centres as well as mergers 
that are concluded by the tidal disruption of the satellite. These two merger criteria are explained in greater  
detail in the method. 

It is necessary to highlight that the simulation used in this work contains dark matter only, while the study of J08 
used a hydro/N-body simulation including gas and star formation. Using a dark matter only simulation has limitations.
 With regard to mergers defined by halo cores coalescing, it is assumed that galaxies lie in the centre of their
 dark matter halos and while this is a reasonable assumption it does not represent a 1-1 mapping when studying effects
 such as AGN activity and SMBH growth. With respect to the tidal disruption merger definition, the disruption of the
 dark matter halo does not necessarily correspond to tidal disruption of the galaxy. It is reasonable to assume that 
there is a connection between the two and at the very least the disruption of the DMH serves as proxy for the
 tidal stripping time of the galaxy.

The aim of this paper is to assess the validity of currently used merger-time formulae with as many mergers
 (of sufficient resolution) as possible. The focus is to better approximate the merger-times of DMHs
 in an astrophysically realistic context even if this does not correspond to an improvement in the description of
 dynamical friction in idealized isolated cases.}

We describe our method in section 2, present
 the results of our comparison in section 3 followed by conclusions in section 4.

\section{METHOD}

We ran a cosmological N-body simulation using the code GADGET2 \citep{gadget2}. We then identify DMHs
using P-groupfinder. For each halo at z=0 we construct a merger-tree, which gives us a detailed dynamical history of
all the halos in our simulation at all redshifts and allows us to look at all the mergers in the volume. Once mergers
are identified we track them in subsequent snapshots on a particle by particle basis. In the following sub-sections we
describe the details of the above steps.

\subsection{Simulation and Group Finding}

A cosmological N-body simulation of dark matter was used to simulate a volume of the universe, 10 $h^{-1}$Mpc$^3$ on one
side, from redshift z=50 to z=0. The relevant cosmological parameters were from WMAP3 data: $\Omega_{\rm M}$=0.24,
 $\Omega_{\rm \Lambda}$=0.76, $\sigma_8$=0.74 and h=0.73 \citep{wmap3}. Our mass resolution is 4.9$\times$10$^5\Msun h^{-1}$ (corresponding to $\rm 512^3$ particles)
 and we saved 102 snapshots between redshifts z=19 and z=0.

The initial conditions (ICs) at z=50 were calculated using the Zel'Dovich approximation, which assumes a 
linear evolution in density from z=1000 to the starting redshift \citep{zeldovich70}. It is true that using a higher
starting redshift and more sophisticated methods of initializing the volume will result is a slightly different mass 
function \citep{crocce06}, we argue that this will not affect the measurement of merger-times for individual mergers. 
 Analysis of $512^3$ runs for 30 and 100 $h^{-1}$Mpc$^3$ volumes (with WMAP 5 and 7 year releases) is currently underway.
The new volumes will be evolved from a redshift greater then 250 and will allow a much more detailed description of the merging populations.

We continue by identifying dark matter halos in a post simulation analysis with P-groupfinder using the
 Friends-Of-Friends (FOF) approach. We used the typical linking length of 0.2 times the mean inter-particle separation
 (corresponding to 3.9 kpc).

\subsection{Merger-Tree}

For every halo at z=0 we created a merger-tree. Our merger-trees were constructed in the conventional fashion whereby
we take a parent halo from snapshot \emph{i} and look for its largest contributor in snapshot \emph{i-1}. Halos are linked to their
 ancestors and descendants by unique particle ids. Some common pitfalls associated with building a merger-tree in
this way are what we refer to as flybys and bridged halos. Flybys occur when the satellite halo enters the host halo only
to continue on a perturbed (but ultimately unbound) trajectory in subsequent snapshots. Bridging happens when
two small satellites appear to merge prior to falling into a much larger potential/halo. This is registered in the merger
tree as two hierarchical mergers, but in reality the larger halo has simply accreted two small satellites.

Both of these problems arise from the discrete nature of simulations outputs. Addressing the impact of these 
systematic errors on galaxy evolution and black hole growth models (e.g. \citealt{volonteri03,somerville08,miro07,miro11,
kauffmann93,croton06}) is beyond the scope of this paper, however further research on this topic is important to understand
 and quantify the consequences this will have on current models ({Poole in prep}, also see \citealt{flybys}). In the appendix
we discuss some of the steps taken in the construction of our merger-trees to avoid “false” or artificial mergers.
 
\subsection{Halo Definitions and Merger-Tree Pruning}

We track halos with at least 500 particles. While the FOF halo mass is used to discern relevant halos in the 
merger-tree, the virial mass and radius are used in all other calculations. We approximate the viral mass and radius by
the spherically averaged over-density. We define a halo where $\rho_{\rm vir}(z)=200\ \rho_{\rm crit}(z)$, in which 
$\rho_{\rm vir}(z)$ is the average density within the virial radius and $\rho_{\rm crit}(z)$ is the critical density of the
 universe at the given redshift. Our merger-tree is constructed using halo masses defined by the FOF-groupfinding algorithm.
 After applying the SO algorithm to these halos we find several of them lack a significant over-density (that would be classified
 as a core). These halos are naturally excluded.

We calculate the centre of each halo potential using an iterative approach (e.g. \citealt{power03}). We calculate the
centre of mass of the group and proceed to remove particles beyond $0.98\ r_i$, where $r_i$ is the current radius ($r_0=r_{\rm vir}$).
 We then re-calculate the centre of mass for new set of particles and repeat the process until the number of particles reaches
$5\%$ of the original set. Halo velocities are approximated using the weighted mean, where weights are based on the distance
from the centre of the potential.

We made several cuts to our collection of mergers in order to make our analysis set. As previously stated, there are a notable number of artificial mergers that we excluded from constructing our
merger-tree. Flybys and bridged mergers have a significant impact on the accuracy of merger-trees, and we developed
measures to exclude these from the analysis (see appendix). In general phase-space merger-tree codes (e.g. \citealt{rockstar}) 
mitigate many of these artificial mergers automatically. Naturally we exclude any merger that does not finish by
the end of our simulation. The effects of this can be seen as a distinctive curve in the upper left corner of the raw data
plots of figure \ref{plot:mtzs}. This introduces a selection bias to our data set which will be discussed later.

\subsection{Defining Merger-Time}

Merger-time can be defined in multiple ways. Most involve the evolution of a characteristic or property of the merger
such as separation, specific angular momentum or the number of bound particles, each has its own pitfall. Separation
becomes an inaccurate proxy for highly radial orbits. Tidal disruption can impede the measurement of kinematics when
using a specific angular momentum criteria. The number of bound particles is a relatively robust condition but is best
tuned to track mergers that entirely disrupt the satellite. It is, however, highly dependent on the criteria for disruption
and the force and mass resolution of the simulation \citep{klypinomerge}.

It is important that we clarify the definition of a merger in its astrophysical context. In this work we use two
 definitions of a merger, however we point the keen reader to \citet{wetzel10}
\footnote{\citet{wetzel10} define 4 types of satellite removal which are directed toward the abundance matching
 approach to galaxy evolution. Semi-analytical and hybrid models should intrinsically incorporate their 3rd and 
4th types of removal \emph{if} flybys and bridged mergers are dealt with in an appropriate fashion by the merger-tree.}. 
The types of merger assessed in this work are: (1) Coalescence, where halo cores (and their hosted
galaxies) coalesce into a new potential before the satellite is tidally disrupted. This is the classically perceived
 picture of mergers in papers such as \citet{tremaine} and \citet{whitedf}. (2) Disruption, where the satellite is tidally
disrupted before the halo cores coalesce. The various stellar streams in our own galaxy \citep{fieldostreams} are
testament to the ability of tidal forces to disrupt satellites before they coalesce with their host.

The start of the merger is defined as the moment the satellite halo crosses the hosts virial radius. To get a 
better approximation of the exact moment we interpolate to first order all merger properties between the adjacent snapshots.
The end of the merger is quantified by two criteria: the evolution of specific angular momentum and the evolution of the
number of bound particles. The Angular Momentum Criteria (AMC) is satisfied when specific angular momentum 
decreases to $5\%$ of its initial value at the start of the merger. The angular momentum approach acts as a good criteria for
coalescence {\myedit and is the same approach taken by BK08. J08s halos have a baryons which can be used to directly track
 exactly when the two galaxies collide}. The Bound Particle Criteria (BPC hereafter) covers tidal disruption.
 Similarly, the BPC is satisfied when the number of bound particles decreases to $5\%$ of the initial value. We
 incorporate both types in a mixed definition, whereby the end of the merger is marked by whichever of the two
 criteria is satisfied first. When we compare our work to the literature we take care to note and match the merger criteria.

\subsection{Tracking Mergers}

We follow a fixed set of particles for the host and satellite halos at the time of in-fall. For the satellite
we continue to use the same set of particles and retrieve their updated positions and velocities in subsequent snapshots.
For the host, in order to take into account accretion, we use the original set of particles to track the central position and
velocity of the halo but re-calculate the mass and radius based on the SO algorithm for all surrounding particles.

This approach eliminates errors typically associated with tracking halos using group-finders. One such error 
occurs when two over-densities of comparable size occupy the same FOF halo. Then the group-finder tracks the largest of
the two as the main halo, but when the over-densities are of comparable size the tag can artificially switch to the other
halo in the next snapshot (see \citealt{wetzel10}). Another problem intrinsic to some subhalo finders (and any technique that
 relies on a density contrast to track substructure) is that they can intermittently lose substructure when it passes by 
the dense centre of the host (see \citealt{ludlow}). The appendix highlights the benefit of using a
fixed particle set to track mergers.

\section{Results}

\begin{figure*}
\begin{center}
\includegraphics [width=\textwidth,angle=0]{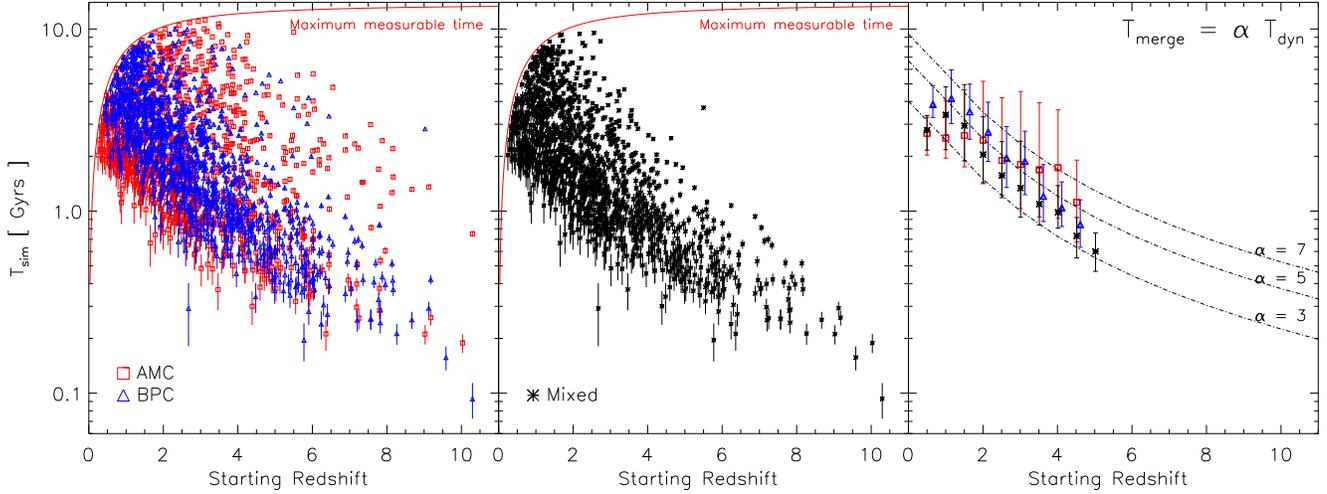}
\caption{Merger-time against starting redshift. The plot holds a total 1438
 mergers with a mixed end definition (threshold 0.05). The left panel shows
the measurements for the AMC and BPC merger sets along with the associated errors.
The middle panel shows the data for the mixed merger definition. The red curve
 for the first two panels represents the maximum
 merger-time measurable at a given starting redshift. The plot on the right
 shows the data for all sets binned in starting redshift.
 The central point for each bin corresponds to the median while error bars
 represent the interquartile range.}
\label{plot:mtzs}
\end{center}
\end{figure*}

Throughout the results section we compare the different types of end-of-merger
 criteria and how they vary with orbital parameters. Recall that in both the AMC and BPC
criteria the merger is considered complete when the value is $5\%$ of its original value at in-fall. When we use AMC only,
our data set consists of 1116 mergers (denoted by red unless stated otherwise). When BPC is applied the data set
consists of 1199 mergers (blue). In the mixed set the end of the merger is defined by whichever criterion is satisfied
first. This gives us 1438 mergers (black). Interestingly, 877 of the mergers have end-time measurements for both AMC
and BPC methods. Furthermore, $55\%$ of this sub-set have $\rm T_{\rm BPC} < T_{\rm AMC}$. Throughout the results section
 any data bin used to compare variables contains a minimum of 40 mergers. The only bins that do not have $\rm N\geq40$ are distribution
functions, usually in the bottom panels. Overall we find that merger-time is most strongly correlated with the dynamical timescale
 of the host halo. With regard to other merger parameters we get varied results for the different merger definitions.

The results are presented in the following form; in section 3.1 we look at the merger-time vs dynamical time 
relation, in section 3.2 we investigate mass ratio, circularity and orbital energy and finally in section 3.3 we compare our measured
merger-times with the predictions of other merger-time formulae and provide our own best fit parameters.

\subsection{Merger-Time with Redshift}

Merger-times for such high redshift halos has remained unexplored in cosmological simulations. Figure \ref{plot:mtzs} shows the
measured merger-time as a function of the redshift at which the merger starts. The left panel shows data from the AMC
and BPC merger sets separately. The central panel shows the mixed data set while the right panel shows the binned
equivalent of each merger set. Bins are spaced equally in starting redshift. Points represent the bin median and error
bars show the inter-quartile range. The red curve shows the limit of our measurements imposed by the end of the 
simulation at redshift z=0 (if the merger does not end in Hubble time, it is excluded from the data set). The
 measurement limits introduced by the end of the simulation will give rise to a selection bias for our low
redshift mergers. This can be seen in the first couple of bins in the right panel in figure \ref{plot:mtzs} as data starts to
 deviate from the fit. In general, all three sets of data show that high redshift mergers take considerably less time to
 finish then those at low redshift

One of the principal components of most prescriptions of merger-time is the dynamical timescale of the host halo
$T_{\rm dyn}$. In the context of halo mergers, dynamical time is the time necessary for a satellite halo to make its first
 pericentric passage of the host halo. The redshift evolution of a halos dynamical timescale in spherical collapse models can take
multiple forms,

\begin{equation}
T_{dyn}=\frac{r_{vir}}{V_c(r_{vir})}=0.1H(z)^{-1}.
\label{eq:tdyn}
\end{equation}

\noindent where $H(z)$ is Hubble's parameter and for the relevant cosmology
 is given by $H^2(z)=H_{0}^2(\Omega_{\Lambda}+\Omega_{m}a^3)$. Note we leave out a factor of $\sqrt{2}$ in equation 2 
as this will later be absorbed in other constants. The growth of a halos mass with redshift enlarges the virial radius
 as well as increases the dynamical time. {\myedit Both BK08 and J08 take this effect into account. BK08 fit their orbital
 parameters to $T_{\rm merge}/T_{\rm dyn}$ (where $T_{\rm dyn}$ is constant for their host) while J08 have a unique 
$r_{vir}/{V_c}$ for each merger}. The right panel in figure \ref{plot:mtzs} shows how the merger-times measured
 in our simulation correlate with dynamical time. We plot 3 multiples of $T_{\rm merge}={\alpha}T_{\rm dyn}$ to illustrate the
 impact of $T_{\rm dyn}$ in our results. We do see indications that there is a redshift dependence beyond the $T_{dyn}$ 
correlation. Such a dependence might manifest in the redshift evolution of the distribution of orbital parameters 
(like \citealt{wetzelorb}). We hope to study this in greater detail with the new simulations.   

\subsection{Other Orbital Parameters}

The dependence of DMH merger-times is often tied to other orbital parameters such as mass ratio,
 circularity of the orbit and the orbital energy. In this section we explore these parameters
 and their dependence on the length of the merger.


\begin{figure}
\includegraphics [width=0.5\textwidth,angle=0]{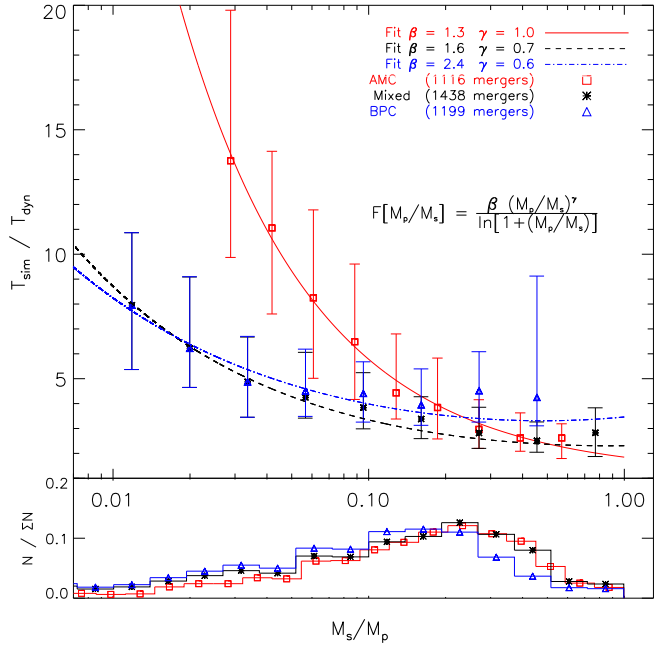}
\caption{A comparison of $T_{\rm merge}/T_{\rm dyn}$ vs mass ratio for the different end criteria.  The central point in each bin represents the median while error bars correspond
 to the interquartile range. Each set is fitted to the equation shown. 
The bottom panel shows the fractional distribution of the merger sets with mass ratio.
The $\chi^2$ fit values are shown in the top right.}
\label{plot:mr}
\end{figure}

\begin{figure}
\begin{center}
\includegraphics [width=0.5\textwidth,angle=0]{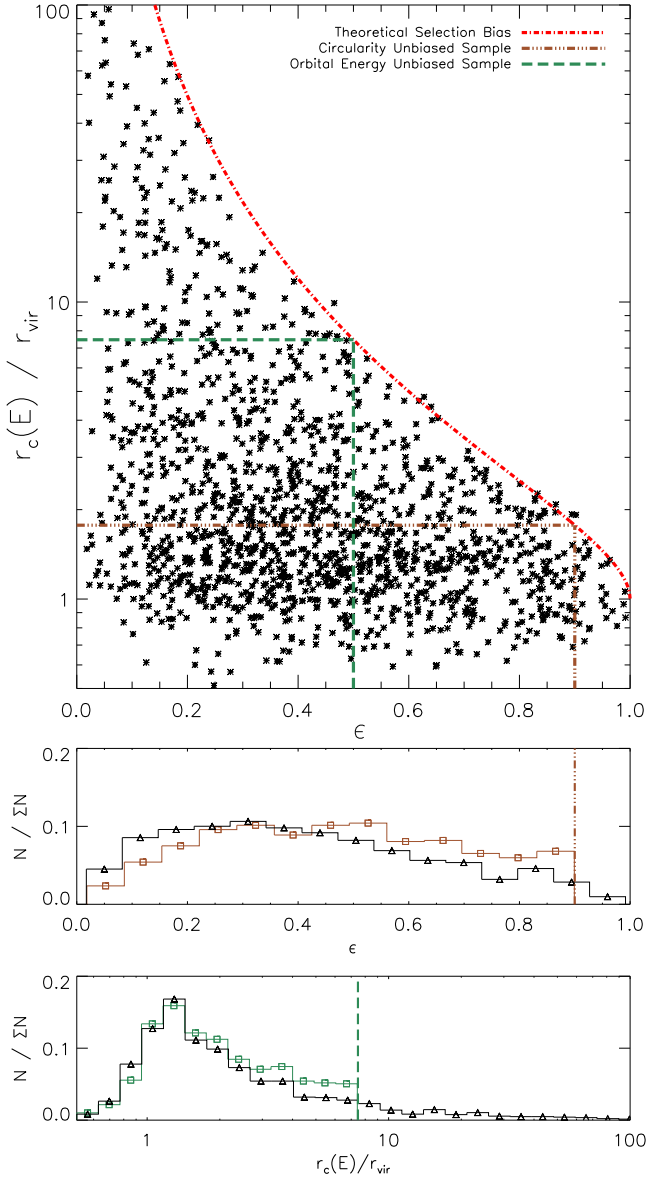}
\caption{Circularity as a function of orbital energy. The top panel shows raw data with the
selection bias (see text) marked by the red line. Also shown in the top panel is the unbiased
 samples used for the frequency distributions in the lower two panels.}
\label{plot:encirc}
\end{center}
\end{figure}

Theory, especially with respect to the classically perceived picture of mergers, expects a rather strong 
dependence of merger-time on mass ratio. There have been a number of works aiming to refine the form of the Coulomb
 logarithm. It is becoming increasingly accepted that the mass ratio dependence of merger-time is best characterized
by:

\begin{equation}
T_{merge } \propto \frac{M_p/M_s}{\rm ln(1+M_p/M_s)},
\label{eq:mr}
\end{equation}

\noindent where the Coulomb logarithm is $\rm ln( 1 + Mp /Ms )$. Figure \ref{plot:mr} shows measured merger-time 
(corrected for the $T_{\rm dyn}$ dependence we see in figure \ref{plot:mtzs}) as a function of the mass ratio of merging halos.
 In all sets of data there is a clear trend that the duration of the merger increases as the host halo becomes larger
 than the satellite. Our results for the AMC mergers are consistent with previous works \citep{wetzel10}.
 While there is considerable scatter, the median value for each bin is in good agreement with the fit.
 The general scatter can be attributed to freedom in other orbital parameters. {\myedit Despite fitting to the functional from
 BK08 (shown in figure \ref{plot:mr}) we find an exact match with J08, i.e that $\gamma=1.0$}.

 When the BPC is used, we see that merger-time does not vary with mass ratio as expected. While the trend is correct,
 the dependence is weaker than seen in other works. The histogram in the bottom panel of
 figure \ref{plot:mr} shows the fractional distribution of merger sets with respect to mass ratio.
 

In any closed two body system the exact trajectory can be described by 
 circularity\footnote{Circularity ($\epsilon$) is related to the eccentricity($e$) by
 $\epsilon=\sqrt{1-e^2}$.} and orbital energy. Both of these variables are hard to constrain
 in such a broad parameter space. {\myedit As seen in J08 it is sometimes easier to look at weaker
 dependences by plotting them against $T_{\rm sim}$/$T_{\rm model}$.}

Figure \ref{plot:encirc} shows orbital energy as a function of circularity for all mergers used 
in our analysis (i.e the mixed set). A selection bias can be seen in the top panel in the form of
 an `envelope' of points. This is the product of the definition of merger start time (i.e the point
 at which the satellite crosses the hosts $r_{\rm vir}$). The detection of mergers with higher orbital energies
 and/or more circular orbits is limited by $r_{\rm per}/r_{\rm vir}\leq 1$; in other words we will only detect
 mergers with orbital pericentres inside the hosts virial radius. An orbits pericentre is defined as
 $r_{\rm per}=a(1-e)$. For circular orbits the semi-major axis $a$ is equal to $r_c$. This
 constraint can be expressed in terms of the circularity of the orbit such that 

\begin{equation}
\frac{r_{per}}{r_{vir}}=\frac{r_c(1-e)}{r_{vir}}\leq 1.
\label{eq:ebias_start}
\end{equation}

\noindent In the limit of this constraint $r_{\rm per}/r_{\rm vir}=1$, equation \ref{eq:ebias_start} can be re-arranged to give,

\begin{equation}
\frac{r_c}{r_{vir}}=\frac{1+\sqrt{1-\epsilon^2}}{\epsilon^2}.
\label{eq:ebias}
\end{equation}

\noindent Equation \ref{eq:ebias} is shown in figure \ref{plot:encirc} as a red curve. It 
is in excellent agreement with the envelope. 

\begin{figure}
\includegraphics [width=0.5\textwidth,angle=0]{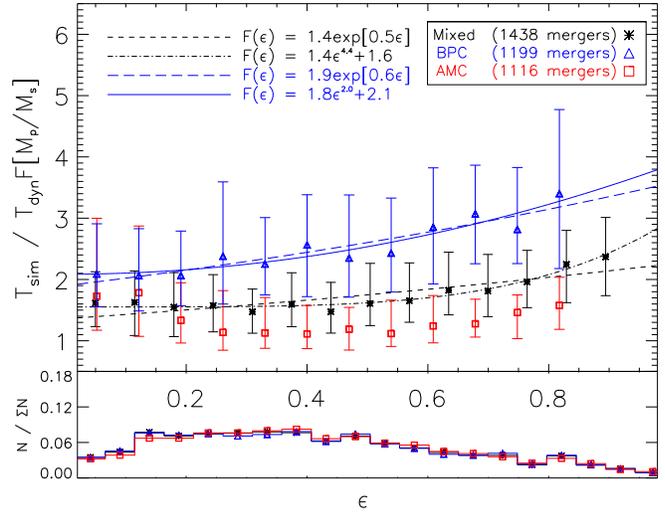}
\caption{Merger-time as a function of circularity. 
Points correspond to the median while errors represent the interquartile range.
The bottom panel shows the fractional distribution of the merger sets with 
respect to circularity. $\chi^2$ fits are shown for the bound and mixed particle sets.}
\label{plot:circ}
\end{figure}

The bottom two panels of figure \ref{plot:encirc} show the distribution of circularity and orbital 
energy respectively. In each panel we show the distribution of the full data set as well as an
 unbiased sample. Sub-samples are set by equation \ref{eq:ebias} and highlighted with lined
boundaries in the top panel. The middle panel shows that selection bias has a strong influence
 on the distribution of circularities. This bias will not influence the fit of merger-time (section 3.3)
 but may have an impact on the fitting of circularity in figure \ref{plot:circ}. The unbiased set is in
 good agreement with \citet{wetzelorb} and \citet{benson05b}.

The dependence of merger-time on circularity is less constrained than the mass ratio relation.
Figure \ref{plot:circ} shows merger-time against circularity for all 3 merger sets, corrected
for the already fitted parameters. The overall trend shows that the duration of mergers increases
 with increasing circularity. 

This positive correlation of merger-time and circularity is seen most strongly in the BPC set. Under the BPC definition,
 the trend corresponds to radial orbits tidally disrupting faster than circular orbits.
 This is expected as radial orbits make close central passings where tidal forces are strongest. 
The AMC merger set deviates from the trend significantly for more radial orbits. This deviation 
persists when an unbiased sample of circularities is used\footnote{It also persists under 
the condition that the satellite has a well resolved core.}. It appears to be physical and not an
artifact of method or numerics. The implication for merger models is that, for well resolved satellites, both
highly radial and highly circular orbits take a similar time to coalesce. Arguably, the binned points
 in figure 9 of J08 show a similar trend. Their trend is much weaker and it would
 require a detailed analysis to confirm it is a physical effect. Radial orbits 
$\epsilon<0.3$ are often overlooked in studies of this type, so perhaps this effect has not been appreciated.

We fit the mixed and BPC merger sets to the dependences seen in J08 and BK08
(shown top left). {\myedit J08 have retained the analytically founded form\footnote{Introducing an additional fitting parameter 
to accomodate the power law for highly radial orbits.} seen in \citet{laceycole} while BK08 use an imperically
 motived exponent. While both forms of the dependence are in accord with the data, the model with an extra fitting
 parameter (J08) shows a tighter agreement.} We are not able to fit the AMC merger set due to the
 upward trend at low circularities. The bottom panel of figure \ref{plot:circ} shows the frequency distribution of each set.

\begin{figure}
\begin{center}
\includegraphics [width=0.5\textwidth,angle=0]{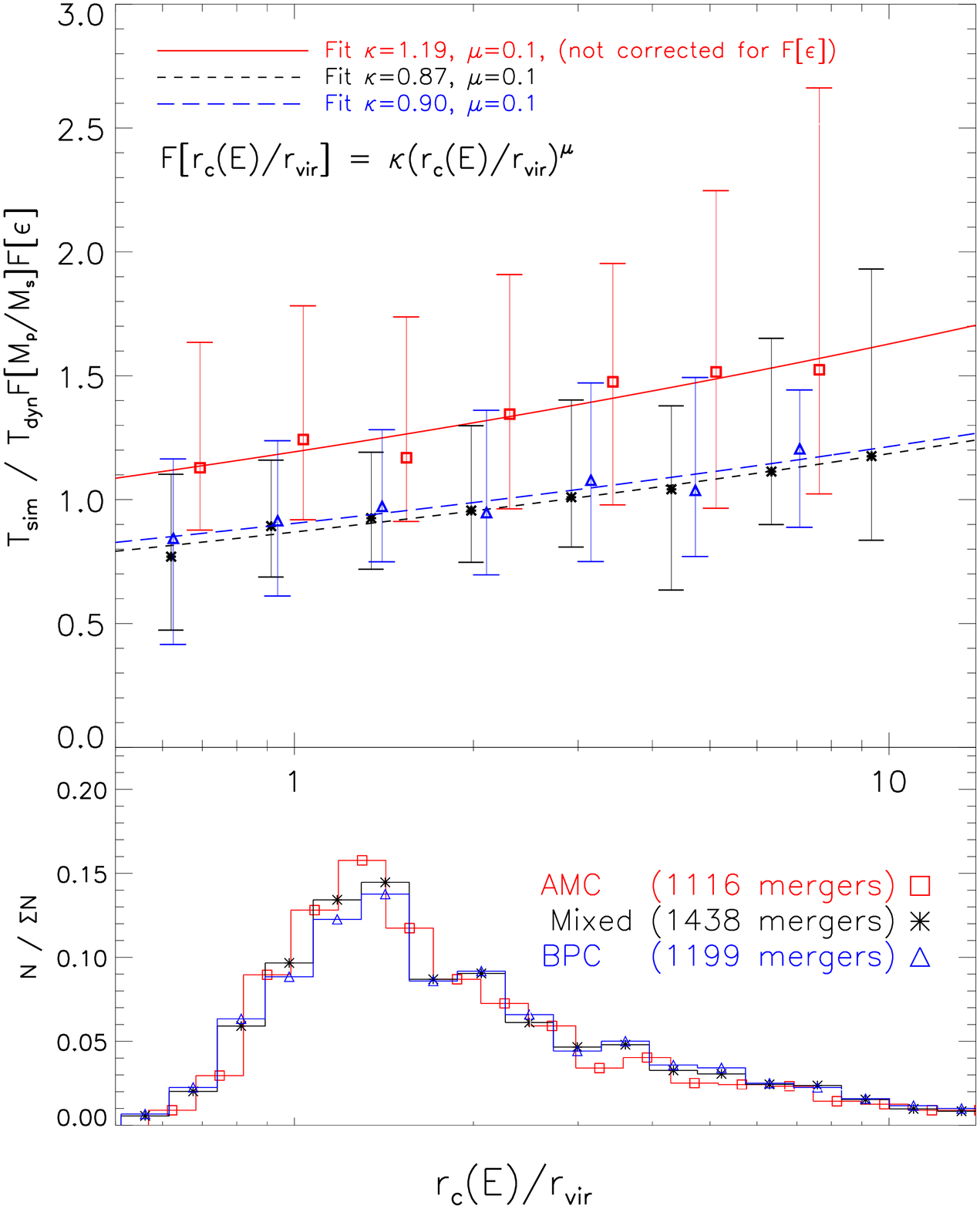}
\caption{The top panel shows merger-time against $r_c(E)/r_{\rm vir}$ (a proxy for orbital energy).
The bins are spaced equally in log. The distribution is skewed towards lower ratios, as shown in the lower 
panel. The $\chi ^2$ fitted parameters are shown
top left. Points correspond to the median while errors show the interquartile range. }
\label{plot:en}
\end{center}
\end{figure}

The final theoretical dependence is the orbital energy of the merger. As previously stated
 the proxy for the energy of the orbit is $r_c(E)/r_{\rm vir}$. This compares the radius of a
 circular orbit (with the same energy as the orbit in question) with the virial radius of 
the host. The top panel of figure \ref{plot:en} compares the dependence of $r_c(E)/r_{\rm vir}$ with merger-time.
Similar to previous figures, other dependencies on merger-time have been taken out. The exception is
the circularity dependence for the AMC set. Overall there is a weak dependence in all 3 data sets.
 Other works suggest $T_{\rm merge}\propto [r_c(E)/r_{\rm vir}]^\mu$, where $\mu\geq1$. {\myedit BK08 find $\mu=1$. 
J08 do not include an orbital energy dependence in their equation 5 for simplicity and ease of use. 
They do however propose an alternate form of the equation in the conclusion that incorperates the
 satellites orbital energy.} We find a relatively consistent value of $\mu=0.1$ across all data sets.  

The fractional distribution of $r_c(E)/r_{\rm vir}$ is displayed in a histogram in the lower panel.
The range of $r_c(E)/r_{\rm vir}$ found in our simulation extends beyond the typically shown range. 
The peak of the distribution is slightly beyond 1. A close examination (snapshot to snapshot) of these
 extremely energetic orbits usually shows a close passing of a third body also orbiting host halo
 (a relatively common occurrence e.g. \citealp{sales07}). 

\subsection{Comparison with previous works}

In this section we compare our measurements with the empirically derived formulae of \citet{boylan}
 and \citet{jiang08}. We remind the reader that the comparison is only with measurements from the AMC merger set as this
is the merger definition under which their formulae were constructed. The best way to make this comparison is to look
 at the variance of the ratio of measured/predicted merger-time ($T_{\rm sim}$/$T_{\rm model}$) for different orbital properties.
 The formulas used in our comparison are 

\begin{center}\textit{\citet{boylan}:}\end{center}
\begin{equation}
T_{merge}=t_{dyn}\frac{A(M_h/M_s)^B}{\rm ln(1+M_h/M_s)}exp\left[C\frac{J}{J_c(E)}\right]\left[\frac{r_c(E)}{r_{vir}}\right]^D,
\label{eq:bk}
\end{equation}

\noindent where $J/J_c(E)\equiv \epsilon$ and their fitted constants are 

\begin{equation}
A=0.216,\   B=1.3,\   C=1.9,\   D=1.0.
\label{eq:bkconstants}
\end{equation}

\begin{center}\textit{\citet{jiang08}:}\end{center}
\begin{equation}
T_{merge}=\frac{f(\epsilon)M_h}{2CM_s\rm \ ln \Lambda }\frac{r_{vir}}{V_c(vir)}
\label{eq:jiang08}
\end{equation}

\noindent where $f(\epsilon)$ and $\rm ln \Lambda$ are the corresponding fitted values from their
 work\footnote{We also compare with the alternate form of their fit (their equation 8) but find
 it to be a worse fit.}, i.e $f(\epsilon)=0.94\epsilon^{0.6}+0.60$ and our eq \ref{eq:mr}.

In addition to the above formulae, we do a maximum likelihood estimate of equation \ref{eq:bk} using
 our merger-time measurements and find the following values.

\begin{equation}
A=0.9,\   B=1.0,\   C=0.6,\   D=0.1.
\label{eq:mle}
\end{equation}

\begin{figure*}
\begin{center}
\includegraphics [width=\textwidth,angle=0]{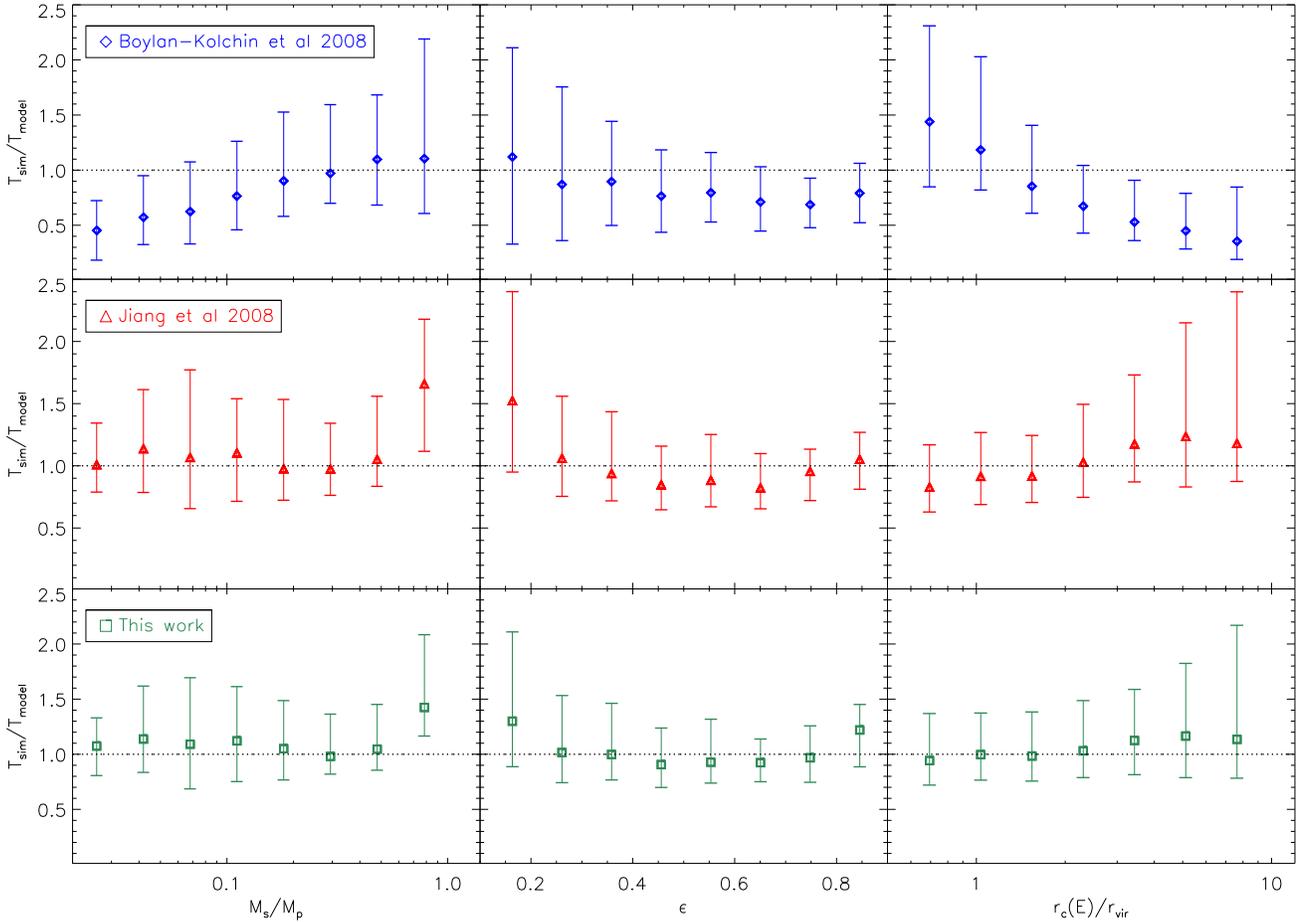}
\caption{ $T_{\rm sim}$/$T_{\rm model}$ against mass ratio, circularity and orbital energy. In this plot we compare the 
empirically derived formulae of BK08 and J08 to the parameters found in this work.
Points correspond to the median while errors show the interquartile range.} 
\label{plot:comparison}
\end{center}
\end{figure*}

\noindent We use these parameters to compare our fit with the other two
works in figure \ref{plot:comparison}. It shows the variance in $T_{\rm sim}$/$T_{\rm model}$ with mass
 ratio, circularity and orbital energy. From top to bottom, figure \ref{plot:comparison} shows a comparison of
 BK08, J08 and the fitted values in this work. BK08 formalism deviates from our
 measurements as a function of mass ratio and orbital energy. At high host-to-satellite mass ratios BK08 overestimate 
the merger-time by a factor of $\sim$ 3. The power law in the mass ratio dependence of BK08 is
 responsible for the deviation. 

\begin{figure}
\begin{center}
\includegraphics [width=0.5\textwidth,angle=0]{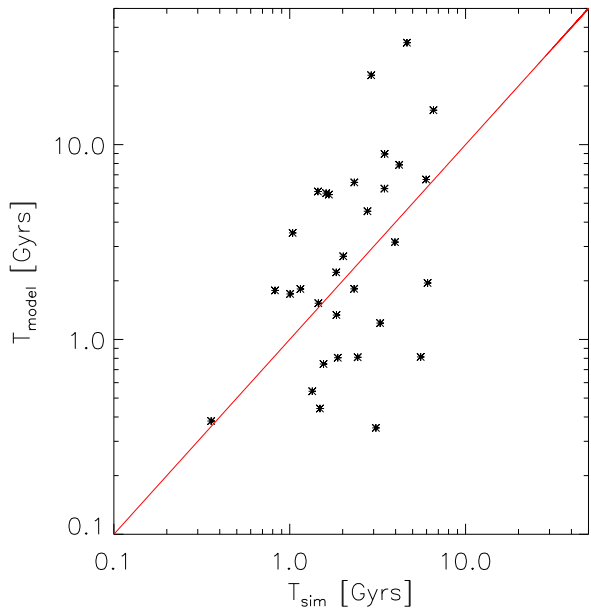}
\caption{A comparison with BK08 showing the predicted and measured merger-times for mergers  
within their valid parameter range.}
\label{plot:bkcomp}
\end{center}
\end{figure}

The discrepancy between BK08 and our fitted parameters (eq \ref{eq:bkconstants} and \ref{eq:mle}) indicates that the dependence of
 merger-time on circularity and orbital energy is weaker than previously derived. This illustrates the difference between
 an idealized isolated merger simulation and a full cosmological simulation. BK08s mergers were in isolation and had
 relaxed Hernquist halos. Conversely, the mergers in this work are in a cosmological context, here the host is still
 collapsing, undergoing multiple mergers at once and at the mercy of the tidal fields in its local environment.

BK08 explicitly state the range of validity of their formula as $0.025 \leq M_s/M_p \leq 0.3$, $0.3\leq \epsilon \leq 1$ and 
$0.65 \leq r_c(E)/r_{\rm vir} \leq 1$ . We find that the range of valid orbital parameters corresponds to a significantly 
limited range of mergers in our simulation. A comparison within this parameter space can be seen in figure \ref{plot:bkcomp}.
It shows a very modest number ($\approx 3 \%$ of our AMC set) of comparable mergers.

The comparison to J08 produces a much tighter agreement in all three parameters
(figure \ref{plot:comparison}, middle panels). Maximum deviation occurs for equal
mass ratio mergers as well as highly radial and energetic orbits. {\myedit Our own fit to BK08s functional 
form shows a slight reduction in scatter when compared with J08. This may just correspond to the different ways
the mergers are qualified. Explicitly that we use specific angular momentum while J08 are able to measure when the 
baryons in the galaxies collide. Either way the results are very well described by the formula of J08.}

As a note, caution should be taken when applying merger-time formulae. The accuracy of a formula can
be highly dependent on the conditions under which it was derived. For example,
 as the merger progresses from some fiducial
 starting point the parameters of the merger change. The mass of the
satellite typically decreases while the mass of the host increases. $r_c(E)/r_{\rm vir}$ 
 tends toward zero, due to the growth of the virial radius with cosmic expansion as
 well as the dissipation of orbital energy from particle-particle interactions. These inputs will
also vary depending on the halo finder/halo definition used to construct the merger-tree.
The application of any approximating merger-time formalism should be consistent with the 
circumstances under which it was derived.

\section{Conclusions}

This paper presents results of a cosmological N-body simulation in a small cosmological volume 
where dominantly spiral and dwarf galaxies form. We studied the mergers
of the dark matter counterparts at high redshift. The mass and redshift range of our simulation
has never been probed for merger-times. Our study considers two merger definitions that qualify the 
end of the merger, tidal disruption and core coalescence. We compare the measured merger-times from our simulation
with the predicted times from other works.

The merger-times of DMHs is strongly dependent on the redshift at which the
merger starts. This manifests itself in the redshift evolution of the host halos 
 dynamical timescale. The dynamical timescale of a halo corresponds (approximately) to the time it takes 
for a point mass to make its first pericentric passage. 

We consider both the AMC and BPC merger definitions (see section 2.5) in our analysis. Both scenarios impact on
satellite removal in N-body/semi-analytic recipes of galaxy evolution. From the mergers that have 
both AMC and BPC end measurements, we find that 55$\%$ have $\rm T_{\rm BPC} < T_{\rm AMC}$. We also find a selection
bias that arises from the pre-requisite that mergers included in our analysis must have $r_{\rm per}/r_{\rm vir}\leq 1$. 
This bias is characterized by equation \ref{eq:ebias} and should be considered in kinematic works of this type.  

In the classical merger definition (AMC - where resolved satellites coalesce) we find good
agreement with $T_{\rm dyn}$ and mass ratio relations. The circularity and orbital energy have less 
of an impact on merger-time. {\myedit Our results show a much weaker dependence on orbital energy than 
analytical/ideal works. We find an interesting result with regard to the circularity 
dependence; for well resolved halos, highly radial and highly circular orbits finish in comparable time.
 This low circularity region ($\epsilon<0.3$) is seldom probed, so it is possible (although unlikely) that
 this effect has been overlooked. BK08 do not cover $\epsilon<0.3$ in their study and J08 do not show as
 strong an upward trend for low circularities. We have taken steps to check for numerical and method
 based artifacts and plan to perform a more in-depth analysis in the future. For $\epsilon>0.3$ we
 find reasonable agreement with other works.}

Our results according to the tidal-disruption merger-criteria (BPC) show some interesting variations.
 We found a tight correlation with the dynamical timescale of the host halo. We also see a dependence on mass ratio,
 although it is weaker then expected (see figure \ref{eq:mr}). Circularity for the BPC set shows shorter
mergers have more radial orbits. In such orbits the satellite is exposed to stronger tidal forces. 

{\myedit We compare our measured merger-times with \citet{boylan} and \citet{jiang08}. We find 
a systematic deviation between our measurements and the predicted times of BK08 who numerically fit their formula 
from isolated mergers. By fitting our cosmological mergers to 
BK08s formula (eq \ref{eq:mle}) we are able to directly compare idealized and cosmological merger-times. As such, we
 find the biggest difference is the dependence on orbital energy ($r_c/r_{vir}$ has a power of 0.1 instead of 1). 	
Our results also show a mass ratio power of 1.0 (formerly 1.3) and a milder circularity dependence. 
We cannot say exactly why there is such a contrast between mergers in a cosomological context and those in isolation, however
part of this discrepancy can be attributed to the fact that in cosmological structure formation 
multiple mergers occur simultaneously. In this work we have empirically found the differences between the two cases  
and the next step is to understand the physical mechanisms responsible. 

In contrast to BK08, our data is well described by J08s approximating formula. Their empirically fitted formula was
 derived under the same cosmological N-body environment as this work. The differences between the simulations in J08
 and this work are the inclusion of baryons as well as the mass and redshift range covered. We remind the reader that there
 are limitations to approximating the AMC and BPC merger measurements using dark matter alone. A comparison of our fit
 with J08 shows a slight reduction in scatter, however this is consistent with the different approaches used to measure
 merger-time. While both formulas fit the data well, the formula of J08 has one less input parameter which makes it
easier to use. Our fit to BK08s form provides slightly reduced scatter at the expense of a additional parameter. 

A comparison of the two works (isolated to cosmological) illustrates that while isolated/idealized 
numerical experiments are important to understand the underlying physics, a more versatile (statistically
robust) approximation may serve better in non-ideal applications, i.e. semi-analytic/hybrid models.}

\section{Acknowledgments}

The authors would like to thank Doug Lin, John Magorrian, Andrew Wetzel, Michael Boylan-Kolchin, Chunyun
Jiang and Greg Poole for helpful discussion and correspondence. We would also like to thank the 
referee for fast and helpful feedback.  
GFL thanks the Australian Research Council for support
through his Future Fellowship (FT100100268) and Discovery
Project (DP110100678).
TM would like to thank the School of Physics, University of Sydney for support 
from the Denison Post-Graduate Conversion Award.

\bibliographystyle{scemnras}

\setlength{\bibhang}{2.0em}
\setlength\labelwidth{0.0em}
\bibliography{bib}

\section{Appendix}

\subsection{Details of our Merger-tree}

\citet{flybys} try to quantify the effect of flybys for corrections in current models. To
 address this issue and avoid ``false mergers'' in the current work we apply the following
 conditions in the construction of our merger-trees.

\begin{itemize}
\item{Any halo associated with the tree has at least 50$\%$ of its particles in the ``root'' halo at z=0.}
\item{If there is any ambiguity as to which halo is the main progenitor (in snap i-1), we then follow
 the contributions of the both progenitors back to i-10.}
\item{One of the benefits of P-groupfinder is that particle ids in a given halo are listed in order of binding
 energy (to a large extent although not exactly\footnote{P-groupfinder lists the particles in order of binding
energy for each substructure. It lists the substructure in a given halo from largest to smallest.})
 this allow us to weight the mutual particles such that more bound particles hold greater weight when considering\
 progenitors \citep{bkweight}.}
\item{When a halo in snapshot i traces to a significantly larger halo in snapshot i-1 (i.e an
 emerging flyby), we look at snapshots i-10 through to i+10 to firstly establish what the halos
 mass was upon in-fall and secondly to see whether the halo grows significantly after it emerges.
 To distinguish between flybys and highly eccentric orbits we employ the condition that a satellite must
at least double its initial in-fall mass to be considered independent after emergence \citep{wetzelorb}.}
\end{itemize}

\begin{figure}
\begin{center}
\includegraphics [width=0.45\textwidth,angle=0]{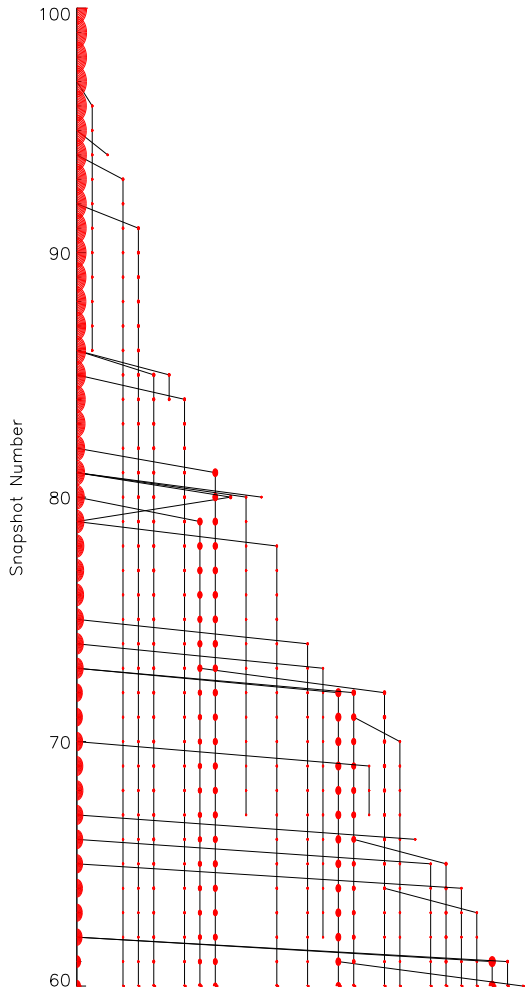}
\caption{An example of one of our merger-trees. The size of the circles is proportional to the size
of the halo.}
\label{plot:tree}
\end{center}
\end{figure}

While these approaches help to minimize the number of flybys it does not catch all of them. 
A small fraction of mergers are false mergers due to flybys. These mergers are removed from
 further kinematic analysis. Figure \ref{plot:tree} shows an example merger-tree.

\subsection{Our Tracking Scheme: The Importance of a Fixed Particle Set}

An initial attempt to track mergers by comparing substructure at different snapshots (via common
particle ids) proved difficult due to the limitations of single epoch group-finders.
 
Subfind identifies substructure by constructing a smoothed 3D density contour of the FOF halo.
 As the global density is lowered, saddle points emerge between adjacent over-densities
 \citep{subfind}. This gives a subset of particles that Poisson's equation can be applied to.

Subfind can temporarily loose substructure when the subhalo is passing through an over-dense
 region. The cause of this intermittent loss arises from the position
based (i.e. smoothed density) selection of particles. It is important to track all
 particles associated with the satellite halo at in-fall when measuring the evolution of 
satellite properties. 

\begin{figure*}
\begin{center}
\includegraphics [width=0.95\textwidth,angle=0]{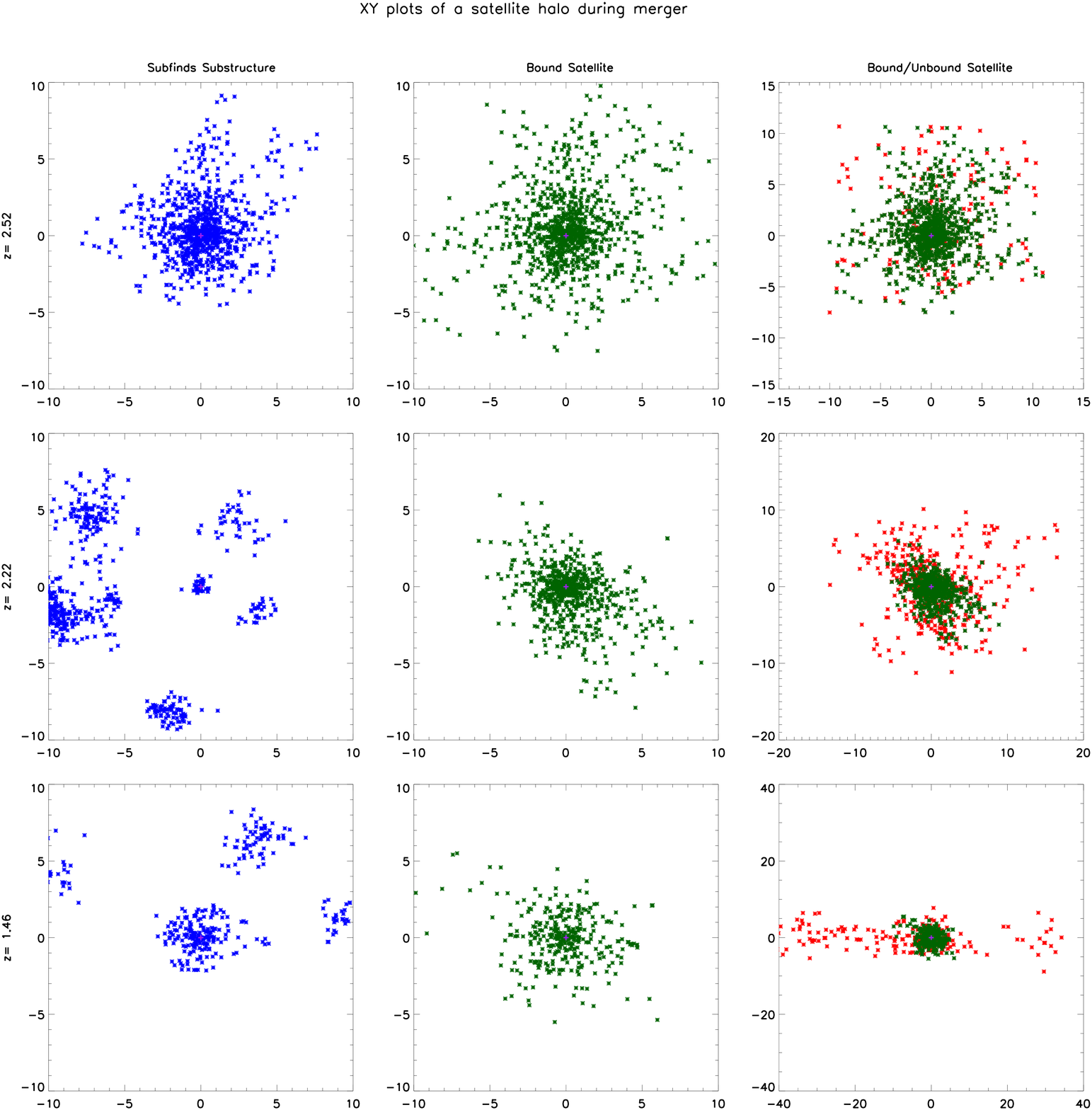}
\caption{A 2D comparison of the structure found with subfind at 3 points during a merger. The
first two columns show the subfind substructure (blue) compared with the satellite tracing
 scheme used in this work (green). The final column shows the bound(green) to unbound (red)
 particles of the satellite in the broader context of tidal stripping. \textbf{Note that column
 3 has a different and varying distance scale to the first two columns}. All columns are 
centred on the same coordinates.}
\label{plot:subfind}
\end{center}
\end{figure*}

Figure \ref{plot:subfind} shows a 2D projection of structure found by various methods at three
 points during the merger. The first two columns show the subfind substructure(blue) compared
 with the satellite tracing scheme used in this work(green). It is important to highlight that 
the blue particles correspond to \emph{all} substructure in the FOF halo and the plots in a 
each row are centred on the same coordinates. The middle row (z=2.22) 
 shows the substructure found as the satellite passes through a region of
 high background density. 

 The third column of figure \ref{plot:subfind} has a different scale to the first two in
 order to highlight the bound particles in reference to their unbound counterparts.
 This demonstrates the effectiveness of our tracking scheme above and beyond the
 groupfinder-particle tracking approach. See \citet{Hanprep} for a more detailed comparison.

\end{document}